# Pressure Induced Superconductivity in $Ba_{0.5}Sr_{0.5}Fe_2As_2$

Georgiy M. Tsoi[1], Walter Malone[2], Walter Uhoya[1], Jonathan E. Mitchell [3], Yogesh K. Vohra[1],

Lowell E. Wenger[1], Athena S. Sefat[3], and S. T . Weir[4]

[1] Department of Physics, University of Alabama at Birmingham (UAB), Birmingham, AL 35294, USA

[2] Department of Physics, University of Portland, Portland OR 97203, USA

[3] Oak Ridge National Laboratory (ORNL), Oak Ridge, TN 37831, USA

[4] Mail Stop L-041, Lawrence Livermore National Laboratory (LLNL), Livermore, CA 94550, USA

## Abstract

High-pressure electrical resistance measurements have been performed on single crystal $Ba_{0.5}Sr_{0.5}Fe_2As_2$ platelets to pressures of 16 GPa and temperatures down to 10 K using designer diamond anvils under quasi-hydrostatic conditions with an insulating steatite pressure medium. The resistance measurements show evidence of pressure-induced superconductivity with an onset transition temperature at ~31 K and zero resistance at ~22 K for a pressure of 3.3 GPa. The transition temperature decreases gradually with increasing in pressure before completely disappearing for pressures above 12 GPa. The present results provide experimental evidence that a solid solution of two 122-type materials, e.g., $Ba_{1-x}Sr_xFe_2As_2$ (0 < x <1), can also exhibit superconductivity under high pressure.

**Introduction**

Discovery of superconductivity in iron-based layered compounds by Hosono and co-workers in 2008 [1, 2] around 30 K has resulted in an explosion of research activity in these types of compounds [1-30]. Undoped 1111-type REOFeAs (RE = trivalent rare earth metal) and 122-type $AFe_2As_2$ (A = divalent elements such as Ba, Sr, Ca and Eu) are non-superconducting materials at ambient pressure and undergo a tetragonal-to-orthorhombic structural transition between 130 K and 200 K upon cooling. This structural transition is accompanied by an antiferromagnetic (AFM) ordering of the Fe at similar temperatures. Both the structural transition and AFM ordering are suppressed under high pressure or chemical doping with the appearance of superconductivity occurring at low temperatures [1-6, 11-13, 17-23, 27-30]. The 122 Fe-based materials have the $ThCr_2Si_2$-type crystal structure at ambient conditions and are of further interest because they undergo a pressure-induced structural transition from a tetragonal-to-collapsed tetragonal crystal structure [26-30] at high pressures. In addition the pressure-induced superconductivity appears to vanish at similar pressures where the collapsed tetragonal crystal structure appears. In addition to pressure-induced effects, superconductivity in 122-compounds can be induced by the chemical substitution of alkali or transition metal ions [3, 11-13] resulting in superconducting transition temperatures as high as ~38 K [3]. For the 122 phase, superconductivity has been induced by substituting Fe with 3d-transition metals such as Co and Ni as well as some 4d- and 5d-transition metals. For example, Ru, Ir, and Pt substitution for Fe have been shown to induce superconductivity in $SrFe_2As_2$ and $BaFe_2As_2$ [20-22]. Superconducting transition temperatures of ~31 K have also been shown to occur by the isovalent substitution of P for As [23]. Thus chemical substitution and pressure provide researchers with an opportunity to tune the structural and magnetic characteristics without changing the nominal charge carrier concentrations in these compounds. Similarly a binary solid solution of the parent $AFe_2As_2$ (A= divalent alkaline earth elements) materials such as $Ba_xSr_{1-x}Fe_2As_2$ ($0 \leq x \leq 1$) may be promising as a new material platform for further investigation of the interplay between the crystal structure, magnetism, and superconductivity under high pressure or chemical substitution or both.

Recent resistivity studies on $Ca_{1-x}Sr_xFe_2As_2$ [32] and $Ba_{1-x}Sr_xFe_2As_2$ [33] samples have reported no evidence of superconductivity at ambient pressure similar to the properties of their parent $AFe_2As_2$ materials. However, the partial chemical substitution of Fe with Co in $Ca_{1-x}Sr_xFe_2As_2$ [32] has led to superconductivity under ambient pressure conditions with maximum superconducting transition temperatures of 17 K. Consequently it would be beneficial to determine if similar pressure-induced structural, magnetic, and superconductivity effects as observed in the parent materials also are present in binary solid solutions at high pressure. In this report, a series of temperature-dependent electrical resistance measurements at high pressure and down to 10 K have been undertaken on $Ba_{0.5}Sr_{0.5}Fe_2As_2$ along with preliminary measurements of the temperature-dependent magnetization and structural characteristics at ambient pressure.

**Experimental Details**

Large platelets of single-crystal, nominal $Ba_{0.5}Sr_{0.5}Fe_2As_2$ were grown from a FeAs flux technique as described in Ref. 11. The crystals were brittle, well-formed platelets with the crystalline [001] direction perpendicular to the plane of the platelets. The platelets were characterized using energy dispersive X-ray spectroscopy (EDS), X-ray diffraction (XRD), temperature-dependent dc-magnetization and electrical resistance measurements. The energy dispersive X-ray spectroscopy (EDS) analyses on several crystals indicated an average Ba/Sr ratio of 52.3: 47.7 for the nominal $Ba_{0.5}Sr_{0.5}Fe_2As_2$ composition. The dc-magnetization as a function of temperature was measured on a single crystal platelet using a Quantum Design model MPMS-5S magnetometer over the temperature range of 5 K to 300 K. The platelet sample was securely mounted between two cardboard disks, which were placed inside a plastic sample holder. By orienting the cardboard disks within the sample holder, the magnetic field was applied parallel to the plane of the platelet (parallel to *ab*-plane) and field-cooled measurements were undertaken at applied magnetic fields of 1.0 kOe and 5.0 kOe. No noticeable difference was observed between the magnetization-vs.-temperature data for the two applied magnetic fields as well as between measurements performed on two different platelets.

The high-pressure and low-temperature electrical resistance measurements were performed down to 10 K and pressures as high 16 GPa using a four-probe designer diamond anvil cell (DAC) technique [34] as described in an earlier publication [16]. Figure 1 (a) shows a picture of the four-probe designer diamond anvil [16] used for the high-pressure resistance measurements. Four tungsten microprobes are encapsulated in a homoepitaxial diamond film and are exposed only near the tip of the diamond to make contact with the sample. Two microprobes are used as electrical leads to send constant current through the sample and two additional microprobes are used to monitor the voltage across the sample as shown in Fig. 1(b). Care was taken to electrically insulate the sample and the designer microprobes from the metallic gasket by using solid steatite as a pressure medium. In addition the solid steatite pressure medium provides for a quasi-hydrostatic pressure measurement. Figure 1(b) shows the $Ba_{0.5}Sr_{0.5}Fe_2As_2$ sample in the insulating steatite pressure medium after finishing a series of electrical resistance measurements at low temperature and high pressure. Pressure was applied using a gas membrane to the designer DAC and the pressure was monitored *in situ* using a ruby fluorescence technique [16, 35] as the DAC was cooled down to 10 K.

**Results and Discussions**

The temperature-dependent magnetization results on a $Ba_{0.5}Sr_{0.5}Fe_2As_2$ platelet at 1.0 kOe are displayed in Fig. 2 along the results measured on platelets of the parent compounds, $BaFe_2As_2$ and $SrFe_2As_2$, under similar conditions. The magnetization measurements at ambient pressure reveal the existence of sharp jumps (indicated by arrows) in the magnetization at 196 K, 164 K, and 131 K for $SrFe_2As_2$, $Ba_{0.5}Sr_{0.5}Fe_2As_2$ and $BaFe_2As_2$, respectively. Since these magnetization jumps in $SrFe_2As_2$ and $BaFe_2As_2$ have been shown previously to result from antiferromagnetic ordering (AFM) of the spin-density-wave (SDW) type, it is reasonable to conclude that the jump measured at 164 K for the $Ba_{0.5}Sr_{0.5}Fe_2As_2$ sample also arises from an AFM ordering of the Fe. Moreover, the AFM ordering for $Ba_{0.5}Sr_{0.5}Fe_2As_2$ occurs at approximately the midpoint of the AFM ordering temperatures of the two parent materials. Lastly, the magnetization results on the $Ba_{0.5}Sr_{0.5}Fe_2As_2$ platelet at ambient pressure show no evidence for superconductivity down to 5 K, similar to the results on the parent compounds.

X-ray diffraction experiments at room temperature and ambient pressure indicate that $Ba_{0.5}Sr_{0.5}Fe_2As_2$ is in the $ThCr_2Si_2$-type tetragonal crystal structure as well, with lattice parameters: $a = 3.94448(8)$ Å and $c = 12.7160(4)$ Å. The calculated volume $V_0$ is 197.847(6) Å$^3$, which is approximately the average of the room-temperature volumes of the two parent compounds. Temperature-dependent structural measurements still have to be undertaken to verify (i) whether a crystallographic phase transition from tetragonal (I4/mmm) to orthorhombic (Fmmm) occurs in this material similar to the parent FeAs materials [3,4,6] and (ii) whether this crystallographic transition is concurrent with the AFM transition at 164 K.

Figure 3 shows the temperature dependence of the electrical resistance of $Ba_{0.5}Sr_{0.5}Fe_2As_2$ at various pressures normalized to the resistance value at 50 K for each applied pressure. These measurements were obtained under nearly hydrostatic condition as insulating steatite was used as the pressure medium. While no evidence for a superconducting resistive transition was found from electrical resistance measurements at very low pressures, a sharp resistive drop to zero resistance is clearly observed at higher pressures, which is characteristic of the superconducting resistive behavior reported on other

AFe$_2$As$_2$ materials. The onset of the resistive transition for the ~3.3 GPa results occurs at $T_c^{onset}$ = 32 K with zero resistance occurring at $T_c^{zero}$ = 22 K. With increasing pressure, the resistive transition gradually shifts to lower temperatures and eventually vanishes for pressures above 13.8 GPa up through the maximum pressure applied of 16 GPa in this study. The evolution of the superconducting transition temperature with pressure is more clearly shown in Fig. 4. The onset temperature $T_c^{onset}$ remains fairly constant for pressures up to 5.6 GPa and then decreases gradually with further increasing pressure up to ~11GPa. The zero resistance temperature $T_c^{zero}$ as a function of pressure behaves qualitatively similar, but typically 8-10 K lower in temperature as compared to $T_c^{onset}$. Experimental evidence for zero resistance disappears for measurements at ~8.8 GPa. Since the present electrical resistance measurements are restricted to temperatures above 10 K, it is possible that the onset and zero resistance transitions occur below 10 K for pressures greater than 11GPa and 8.8 GPa, respectively. Extrapolation of $T_c^{onset}$ to zero temperature would indicate that superconductivity is completely suppressed above ~12.5 GPa, which is close to the pressure for which Ba$_{0.5}$Sr$_{0.5}$Fe$_2$As$_2$ is expected to transform from a tetragonal to collapsed tetragonal phase as described later in this section.

The superconductivity in Fe-based compounds is known to be quite sensitive to structural distortions and the degree of non-hydrostaticity of the applied pressure [17-19]. An increasing uniaxial pressure component has been shown to suppress the antiferromagnetic and orthorhombic phase transitions and to favor the appearance of superconductivity at lower pressures in undoped BaFe$_2$As$_2$ and SrFe$_2$As$_2$ [17-18]. Previous studies using designer DAC performed without any pressure medium typically resulted in broad and non-zero resistive transitions at low temperatures [16, 27-29, 31]. Figure 5 displays the temperature-dependent resistance results on a Ba$_{0.5}$Sr$_{0.5}$Fe$_2$As$_2$ sample taken from the same batch as that used in the "quasi-hydrostatic" experiments, but without any pressure medium and thereby creating a highly non-hydrostatic pressure condition on the sample. At 4 GPa, a resistive transition characterized by broad downturn is observed, but the normalized resistance remains finite even down to 10 K. The resistive transition is suppressed by ~7 GPa and further increases in pressure up to 24 GPa lead to an increase in resistance below 50 K. This is in sharp contrast to the present quasi-hydrostatic measurements using solid steatite as a pressure medium where a zero-resistance state is achieved in the pressure range of 3.3 to 8.8 GPa and the onset of the resistive transition persisted up to ~11GPa.

Figure 6 depicts the linear relationship between the pressure $P_{T-CT}$ resulting in the tetragonal-to-collapse tetragonal phase transition at room temperature and the ambient pressure volumes $V_0$ for the four parent 122-Fe-based materials, $AFe_2As_2$ where A=Ba, Sr, Eu and Ca. Based on the ambient volume of $Ba_{0.5}Sr_{0.5}Fe_2As_2$ determined from our room-temperature XRD measurements and this linear fit, the corresponding pressure for its tetragonal-to-collapsed tetragonal transition should occur at ~13.5 GPa. It is worthwhile to note that this pressure $P_{T-CT}$ is comparable to the extrapolated pressure in Fig. 4 where the onset of superconductivity vanishes. This correlation in applied pressures between the vanishing of the pressure-induced superconductivity and the appearance of the collapsed tetragonal structural phase is in qualitative agreement with prior pressure-induced studies on $SrFe_2As_2$ and $BaFe_2As_2$ and suggests that superconductivity may only be stable in the non-collapsed tetragonal structure [27-29]. Further structural studies are necessary to fully understand this interconnection between the appearance/disappearance of superconductivity and the structural properties of the 122 materials under pressure.

**Conclusions**

In this paper, pressure and temperature-dependent electrical measurements are reported on $Ba_{0.5}Sr_{0.5}Fe_2As_2$ using a designer DAC to pressures up to 16 GPa and temperatures down to 10 K. The resistance measurements show an evidence of a pressure-induced superconductivity with a maximum $T_c^{onset}$ of 32 K at 5.6 GPa. The onset transition temperature decreases with increasing pressure and disappears completely between 10 and 12.5 GPa. The present study provides experimental evidence that pressure can be used to induce superconductivity in a binary solid solution of two 1-2-2 type materials, $Ba_{1-x}Sr_xFe_2As_2$. The correlation between the extrapolated applied pressure resulting in the vanishing of the pressure-induced superconductivity and the interpolated pressure for the structural transformation from a tetragonal-to-collapsed tetragonal phase at 13.5 GPa suggests that superconductivity may only stable in the non-collapsed tetragonal structure, in agreement with previous results on $SrFe_2As_2$ and $BaFe_2As_2$.

**Acknowledgment**

Walter Malone acknowledges support from the National Science Foundation (NSF) Research Experiences for Undergraduates (REU)-site under grant no. NSF-DMR-1058974 awarded to UAB. Walter Uhoya acknowledges support from the Carnegie/Department of Energy (DOE) Alliance Center (CDAC) under grant no. DE-FC52-08NA28554. Research at ORNL is sponsored by the Division of Materials Sciences and Engineering, Office of Basic Energy Sciences, US Department of Energy.

**Figure Captions**

**Figure 1**. (a) Picture of the four-probe designer diamond anvil used in the high-pressure electrical resistance measurements. (b) The $Ba_{0.5}Sr_{0.5}Fe_2As_2$ sample in the insulating steatite pressure medium.

**Figure 2.** The field-cooled magnetization ($M/H$) as a function of temperature for $Ba_{0.5}Sr_{0.5}Fe_2As_2$ in an applied magnetic field of 1.0 kOe. The $M/H$-*vs.*-$T$ results for the parent compounds, $Ba_2Sr_2Fe_2As_2$ and $Ba_2Sr_2Fe_2As_2$, measured under same conditions are included for comparison. The arrows indicate the temperature of the antiferromagnetic ordering for these three compounds.

**Figure 3.** The temperature dependence of the electrical resistance of $Ba_{0.5}Sr_{0.5}Fe_2As_2$ at various pressures normalized to the resistance value at 50 K. Insulating steatite was used a pressure medium.

**Figure 4.** The pressure dependence of the onset temperature $T_c^{onset}$ of the resistive transition and the temperature $T_c^{zero}$ where zero resistance was measured.

**Figure 5.** The temperature dependence of the electrical resistance of $Ba_{0.5}Sr_{0.5}Fe_2As_2$ at various pressures normalized to the resistance value at 50 K when no pressure medium was used.

**Figure 6.** The pressure $P_{T-CT}$ resulting in the tetragonal-to-collapse tetragonal phase transition at room temperature as a function of the ambient pressure volumes $V_0$ for the four parent 122-Fe-based materials, $AFe_2As_2$ where A= Ca [26], Eu [27], Ba [28] and Sr [29, 30]. The solid line represents a linear fit to the experimental data (red solid circles). The corresponding pressure value $P_{T-CT}$ for $Ba_{0.5}Sr_{0.5}Fe_{e2}As_2$ (blue solid square) is determined from this fit using its measured ambient pressure volume $V_0$ as described in the text.

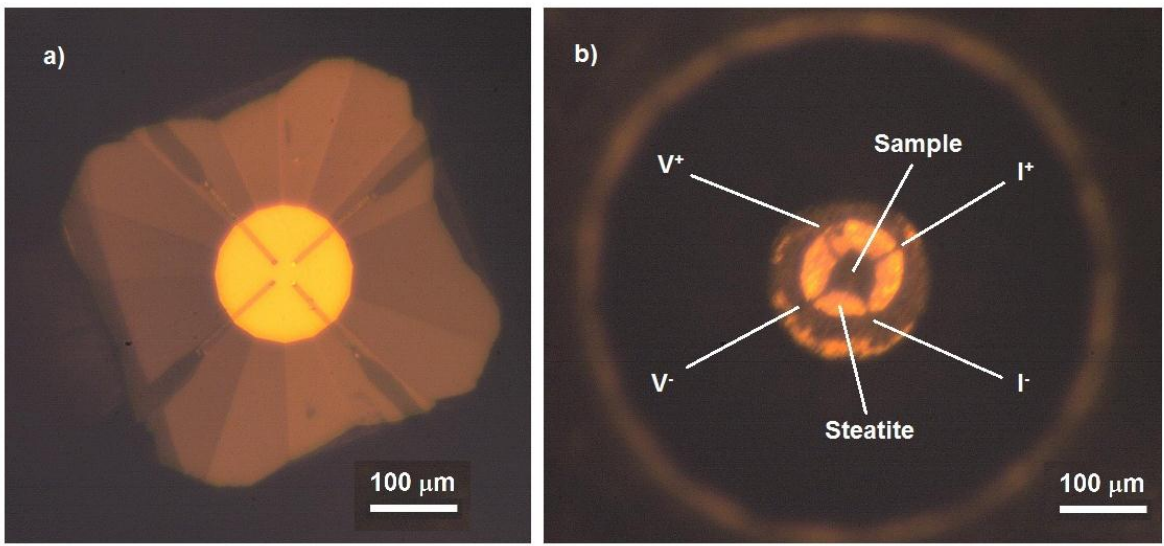

Figure 1

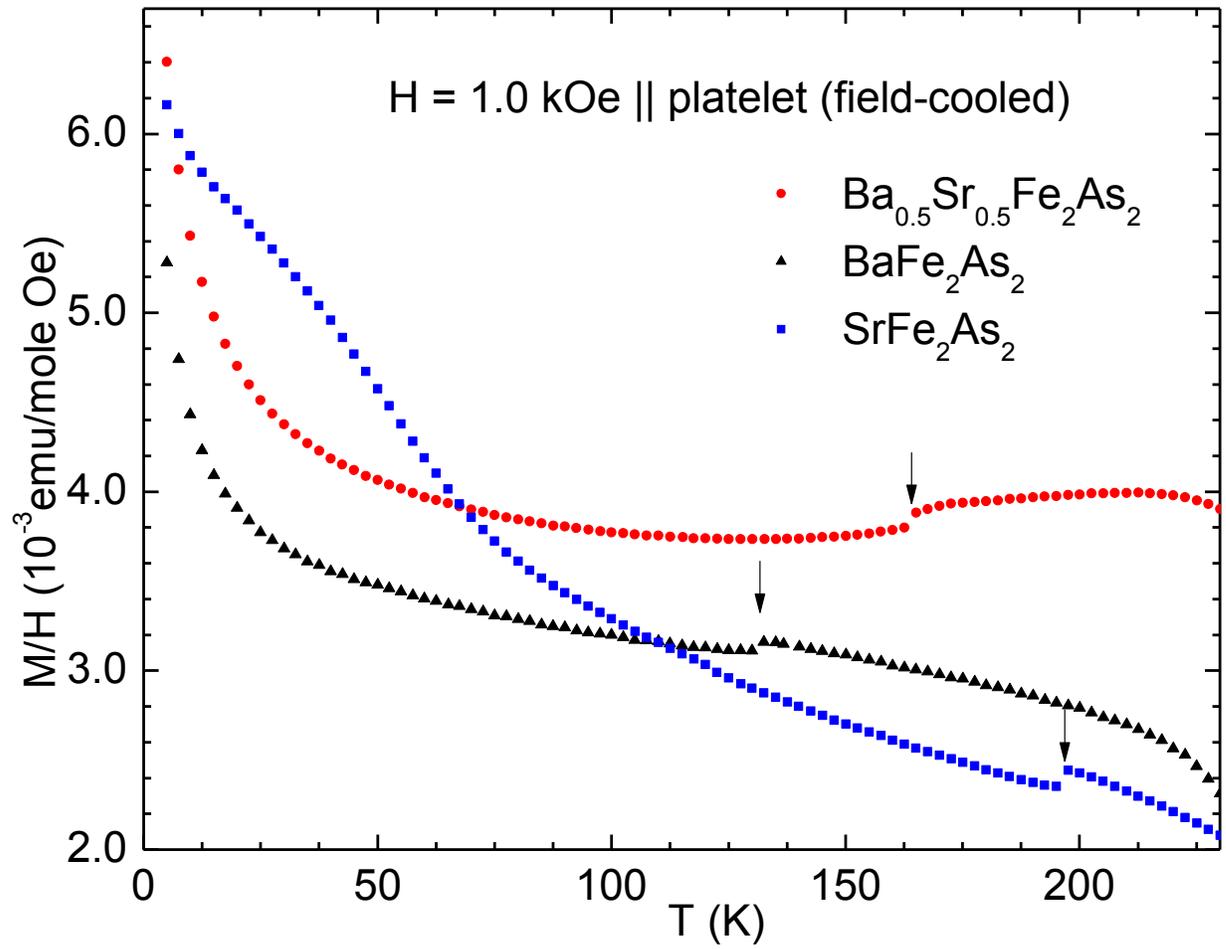

H = 1.0 kOe || platelet (field-cooled)

- $Ba_{0.5}Sr_{0.5}Fe_2As_2$
- $BaFe_2As_2$
- $SrFe_2As_2$

Figure 2

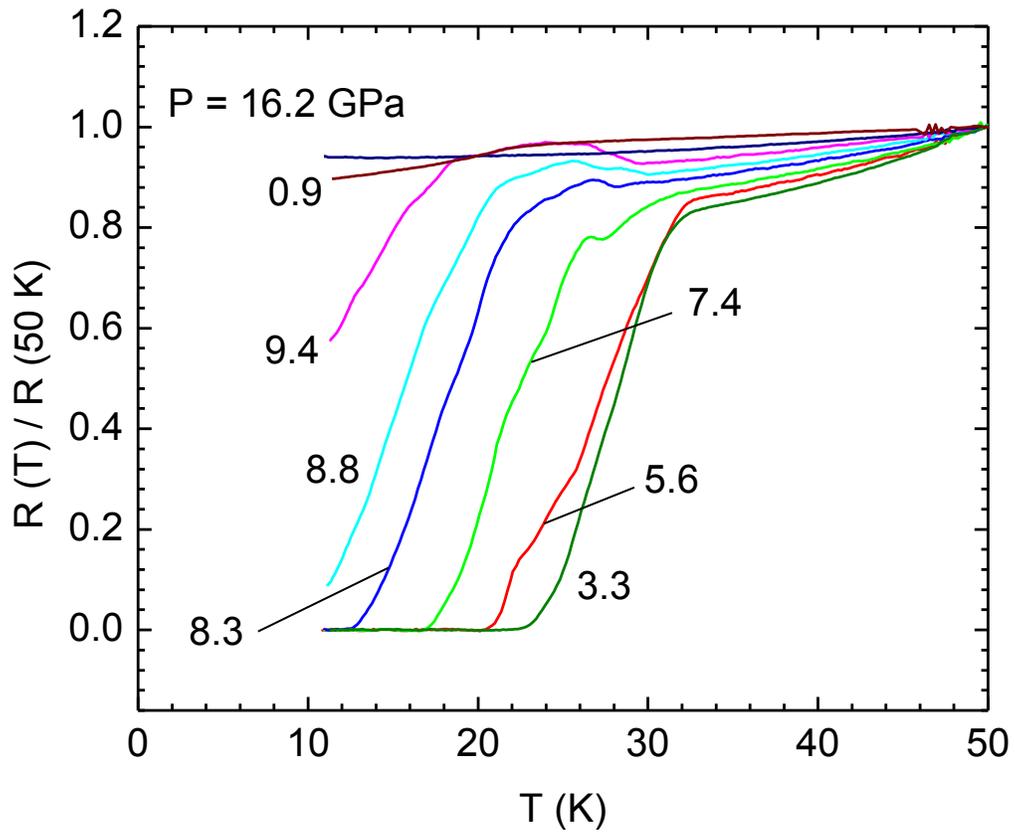

Figure 3

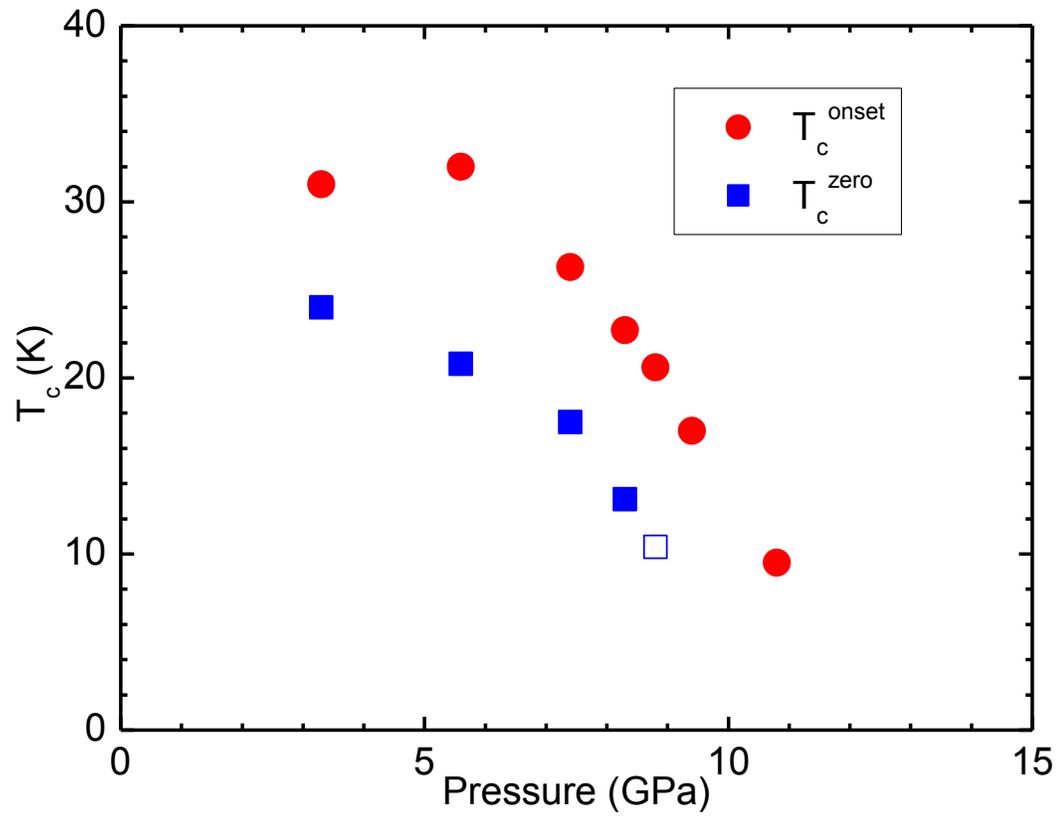

Figure 4

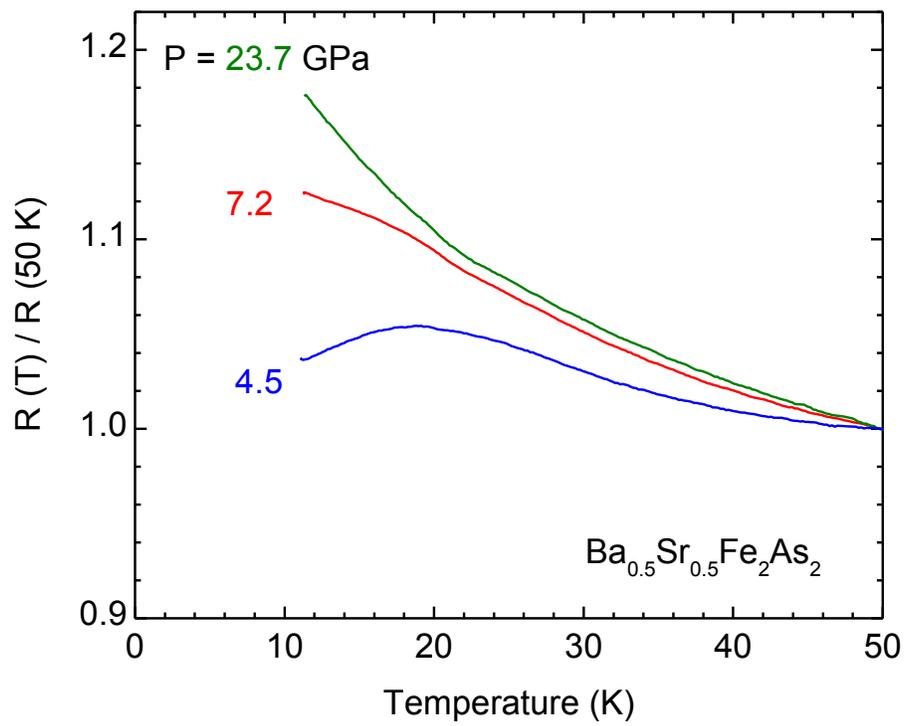

Figure 5

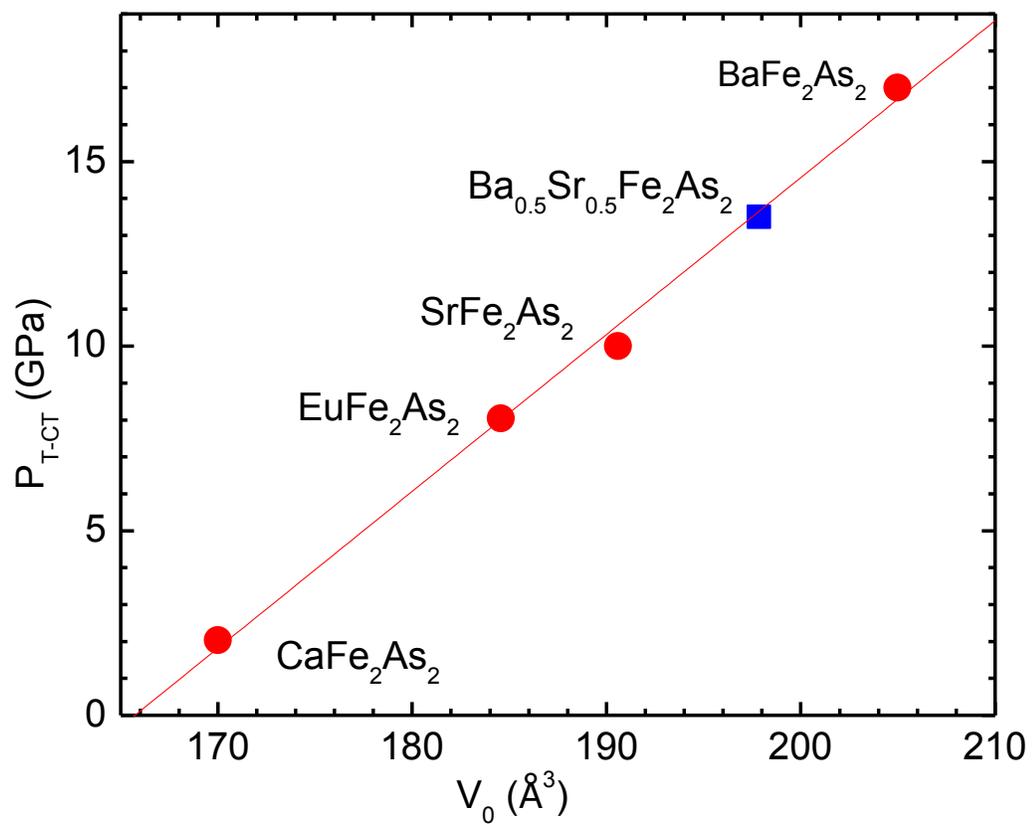

Figure 6